\documentclass{article}

\usepackage[final]{neurips_2026}

\usepackage[utf8]{inputenc}
\usepackage[T1]{fontenc}
\usepackage{hyperref}
\usepackage{url}
\usepackage{booktabs}
\usepackage{amsfonts}
\usepackage{amsmath}
\usepackage[capitalize,nameinlink]{cleveref}
\usepackage{nicefrac}
\usepackage{microtype}
\usepackage{xcolor}
\usepackage{graphicx}
\usepackage{tikz}
\usetikzlibrary{positioning,arrows.meta,fit,backgrounds,calc}
\usepackage{listings}
\usepackage{enumitem}
\setlist{topsep=2pt, itemsep=2pt, parsep=0pt}

\AtBeginDocument{%
  \setlength{\abovedisplayskip}{5pt plus 2pt}%
  \setlength{\belowdisplayskip}{5pt plus 2pt}%
  \setlength{\abovedisplayshortskip}{3pt plus 2pt}%
  \setlength{\belowdisplayshortskip}{3pt plus 2pt}%
}

\makeatletter
\renewcommand{\@noticestring}{}
\makeatother

\lstdefinestyle{dsl}{
  language=Python,
  basicstyle=\scriptsize\ttfamily,
  keywordstyle=\bfseries,
  commentstyle=\itshape\color{black!60},
  columns=fullflexible,
  keepspaces=true,
  frame=single,
  framerule=0.3pt,
  xleftmargin=4pt,
  xrightmargin=4pt,
  aboveskip=6pt,
  belowskip=4pt,
  title={},
}
\lstdefinestyle{ir}{
  basicstyle=\scriptsize\ttfamily,
  keywordstyle=\bfseries,
  morekeywords={Op,OpParams,InnerLoop,GraphParams,LeafParams,Union,Graph,List,Tensor,OpCost,RRT},
  columns=fullflexible,
  keepspaces=true,
  frame=single,
  framerule=0.3pt,
  xleftmargin=4pt,
  xrightmargin=4pt,
  aboveskip=6pt,
  belowskip=4pt,
}

\title{Design Docs Are All You Need: \\
  An AI-native Machine-Learning Performance Tool}

\author{%
  \hspace{-1cm}
  \textbf{Samuel Kushnir}$^1$ \quad
  \textbf{Kimia Noorbakhsh}$^2$ \quad
  \textbf{Kavya Sreedhar}$^4$ \quad
  \textbf{Liqun Cheng}$^4$ \quad
  \textbf{Ming Liu}$^4$ \\
  \hspace{-0.5cm}
  \textbf{Parthasarathy Ranganathan}$^4$ \quad
  \textbf{Mohammad Alizadeh}$^2$ \quad
  \textbf{Fred Kjolstad}$^3$ \quad
  \textbf{Suvinay Subramanian}$^1$ \quad
  \\
  \vspace{0.2cm}
  $^1$Google DeepMind \quad
  $^2$MIT \quad
  $^3$Stanford \quad
  $^4$Google
}

\begin{document}

\maketitle
\begin{abstract}
Machine-learning performance modeling is a uniquely hostile terrain for long-lived
software: the assumptions baked into today's abstractions are invalidated
by tomorrow's models and systems, forcing perpetual refactoring of performance-
modeling frameworks. Meanwhile, AI coding agents have become fast and capable
enough that regenerating an entire library is cheaper than paying down the tech
debt of incrementally patching it. We describe \textsc{smart}, a rigorous
symbolic performance-modeling library for ML systems whose \emph{main branch
contains almost no code}: the repository is a DAG of self-contained natural-
language design docs, coding sub-agents regenerate the implementation from only the
docs on new version updates, and every human change is a
natural-language edit to a doc---self-documenting by construction. Two
ingredients make regeneration reliable: (i) a design-doc style built around
step-by-step worked examples that act as in-context demonstrations for the
generating agents, and (ii) a minimal, recursively
defined operator IR with symbolic (SymPy) cost expressions, a fast analytical
roll-up mode for large sweeps, and a slow modulo-scheduling mode for
fine-grained schedule studies. Regenerated implementations reproduce hand-audited
reference models---including DeepSeek-V3 serving on a TPU pod slice---to
round-off precision, suggesting that design docs---not code---can be the
durable artifact for ML-systems co-design tools.
\end{abstract}

\section{Introduction}
\label{sec:intro}

Machine-learning performance modeling is a punishing environment for software
abstractions, because the assumptions of current architectures and
systems are constantly evolving. An abstraction that seemed reasonable a year
ago---for instance, that every transformer layer looks the same---is suddenly
false in the face of mixture-of-experts routing, latent attention, and
heterogeneous inference phases. Performance-modeling frameworks sit at the
intersection of the two fastest-moving parts of the stack—rapidly evolving model architectures on top, and underlying hardware accelerators and interconnects on the bottom—so they absorb this
churn from both directions; the result is a never-ending stream of
refactorings and hacks to absorb each new architecture or system change into
an aging framework.

At the same time, AI coding tools have become remarkably capable and fast.
Together, these two forces produce failure modes that compound technical debt:

\begin{enumerate}
\item \textbf{Incremental generation debt (predates AI).} Let $S_t$ denote the
specification of a library at time $t$ and let $G$ denote a code generator---a
human engineer or an agent. A greenfield build computes $C_t = G(S_t)$. In
practice, however, the generator at time $t{+}1$ is handed the previous
implementation as an extra argument and computes
$C_{t+1} = G\!\left(S_{t+1},\, C_t\right)$. We can define the technical debt of
this process as
\begin{equation}
D_{t+1} \;=\; \bigl\lVert\, G(S_{t+1},\, C_t) \;-\; G(S_{t+1}) \,\bigr\rVert,
\label{eq:debt}
\end{equation}
the distance between the incrementally patched system and the one that would
have been built from the current spec alone. Because starting over---deleting
all of one's code---is mentally challenging and time-consuming for humans,
the technical debt is in practice much larger than zero: incremental patches inevitably introduce compromises that compound with every spec revision.
\item \textbf{Context-window myopia} AI coding agents are constrained by finite context windows, making it impossible to pass an entire mature codebase into a single prompt. Consequently, developers must feed the agent fragmented, localized code snippets when requesting revisions. Because the agent cannot see the whole picture—missing global invariants, cross-module dependencies, and the broader architectural intent—it frequently generates code that is locally plausible but globally sub-optimal. Over time, these piecemeal, context-blind updates degrade the structural coherence of the framework.
\end{enumerate}

\begin{figure}[t]
\centering
\begin{tikzpicture}[
  scale=0.72, transform shape,
  font=\small,
  doc/.style={draw, rounded corners=1pt, fill=blue!8, minimum width=8mm,
              minimum height=5mm, inner sep=2pt},
  stage/.style={draw, rounded corners=2pt, align=center, minimum height=11mm,
                inner sep=5pt},
  arr/.style={-{Stealth[length=2.2mm]}, thick},
  lbl/.style={font=\scriptsize\itshape, align=center},
]
\node[doc] (d1) {\scriptsize doc};
\node[doc, below left=4mm and 1mm of d1] (d2) {\scriptsize doc};
\node[doc, below right=4mm and 1mm of d1] (d3) {\scriptsize doc};
\node[doc, below=4mm of d2] (d4) {\scriptsize doc};
\node[doc, below=4mm of d3] (d5) {\scriptsize doc};
\draw[arr] (d1) -- (d2);
\draw[arr] (d1) -- (d3);
\draw[arr] (d2) -- (d4);
\draw[arr] (d3) -- (d5);
\draw[arr] (d2) -- (d5);
\begin{scope}[on background layer]
\node[draw, dashed, rounded corners=2pt, fit=(d1)(d2)(d3)(d4)(d5),
      inner sep=3.5pt, fill=black!2] (dag) {};
\end{scope}
\node[lbl, below=1.5mm of dag] {design-doc DAG\\(checked into \texttt{master})};
\node[stage, right=13mm of dag, fill=orange!10] (agents)
  {agentic regeneration\\ \scriptsize one sub-agent per doc,\\ \scriptsize topological order};
\node[stage, right=13mm of agents, fill=green!8] (code)
  {generated library\\ \scriptsize symbolic cost model\\ \scriptsize (build product)};
\node[stage, below=7mm of code, fill=red!6] (valid)
  {validation\\ \scriptsize reference-model reconciliation,\\ \scriptsize parameter guards, unit tests};
\draw[arr] (dag) -- (agents);
\draw[arr] (agents) -- (code);
\draw[arr] (code) -- (valid);
\draw[arr] (valid.west) -| node[lbl, pos=0.25, below] {repair} (agents.south);
\draw[arr] ($(dag.north)+(0mm,6mm)$) -- node[lbl, right=1mm, pos=0.45]
  {human edits: natural language only} (dag.north);
\end{tikzpicture}
\caption{The regeneration workflow: sub-agents regenerate the implementation
from the design-doc DAG in dependency order, and the result must pass
reconciliation against hand-built references before it replaces the previous
build. Humans only ever edit the docs.}
\label{fig:workflow}
\end{figure}
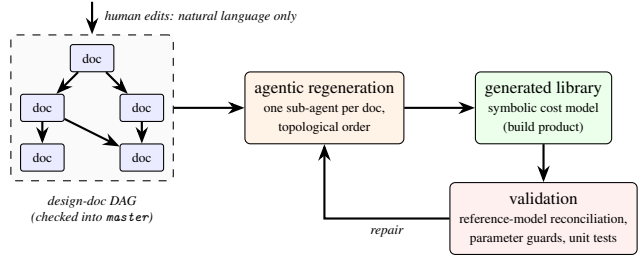

To solve these problems while exploiting the speed of AI coding agents, we
make natural-language \emph{design docs} the durable artifact and treat code
as a regenerable build product (\Cref{sec:docs}). Reliable regeneration
imposes real demands on how docs are written (\Cref{sec:writing}) and how the
modeled system is factored into abstractions (\Cref{sec:ir}). Because $C_t = G(S_t)$ is recomputed from scratch
on a regular cadence, \Cref{eq:debt} is driven to zero \emph{by
construction}. This complete regeneration workflow is illustrated in \Cref{fig:workflow}.

\section{Design docs as the source of truth}
\label{sec:docs}
Our main branch contains almost no code. Instead, it consists of a folder
structure of design document markdown files that compose into a directed
acyclic graph (DAG). Some modules must be generated before others---for
example, hardware topology and numerics precede the collective-cost models,
which in turn precede the model catalog. 

The edges of this DAG are machine-discovered rather than hand-maintained. We
utilize a distributed process where read-only agents analyze the design
documents and infer the dependency edges between them. Once the graph is
resolved, an orchestrator agent walks the DAG in topological order,
assigning a dedicated coding sub-agent to implement each self-contained doc.
Throughout this process, the orchestrator maintains a central log file
recording areas where sub-agents struggled to interpret the prose, as well
as bugs uncovered in the output of agents from earlier topological waves.

Factoring the codebase into self-contained documents and deploying a distinct
sub-agent for each yields three distinct benefits:

\begin{enumerate}
\item \textbf{Bounded context windows.} By restricting the scope of each
generation step to a single document, we limit the context window and task
length for each individual LLM. This avoids overwhelming the model,
significantly increasing the probability of correct code generation for each
isolated task.
\item \textbf{Targeted human iteration.} The orchestrator's log provides
fine-grained visibility into the ``hardness'' of each design doc. By
observing where agents struggle the most and introduce the most bugs, human
engineers know exactly which documents require prose refinement and clarification
for the next version.
\item \textbf{Dynamic model routing.} The orchestrator can dynamically
allocate varying levels of intelligence based on document complexity. While
foundational documents (such as the core DSL design) might be routed to a
larger, more capable LLM, many downstream design docs can be successfully
translated into working code using smaller, more cost-effective models.
\end{enumerate}

\paragraph{Cost and iteration speed.}
In practice, a full clean-slate regeneration of the library takes between 1.5
and 3 hours. The API cost for a complete rebuild using Claude Code is around 100 USD and under a standard high-tier usage plan (e.g., Claude Max), this represents
roughly 20\% of a weekly usage budget, making continuous, full-library
regeneration both practical and economically viable.

\paragraph{An opinionated way of writing design docs}
\label{sec:writing}

Conventionally, in the age of AI, practitioners have focused on writing the
right tests and high-level rules to ensure code correctness and adherence to
project conventions. This is, in essence, a top-down approach---a
\emph{constitution} the generator must obey. We believe a complementary,
bottom-up ingredient is crucial: \textbf{worked examples}. Writing out how a
piece of pseudo-code executes on a given input, step by step---intermediate
shapes, intermediate values, the exact closed-form cost expression that should
result---is what maintains consistency across independent agentic
code-generations. Like a human learner, in-context learning benefits from a
tangible walk-through~\citep{brown2020language,dong2022survey}: a concrete trace pins down
semantics that prose alone leaves ambiguous, directly attacking failure mode 2
of \Cref{sec:intro}. Our docs therefore favor executable-in-your-head vignettes
(``on a $2{\times}2{\times}2$ torus with wraparound, the per-node link count is
3, not 6; the all-gather of $V$ bytes therefore costs \ldots'') and every number-bearing doc ends with a reconciliation anchor: a small preset whose expected outputs are stated exactly and enforced
by generated tests.

\section{A minimal symbolic IR for performance co-design}
\label{sec:ir}

Reliable regeneration also constrains the \emph{artifact being specified}: the
abstractions must be few, orthogonal, and stable under architecture churn. We
use a flexible and minimal IR with both a \emph{fast mode} for large sweeps
and a \emph{slow mode} for careful scheduling studies.

\paragraph{The \texttt{Op} abstraction.} A model is defined by
a single recursively defined operation:

\begin{lstlisting}[style=ir]
Op:
  inputs:  List[Tensor]     # symbolic shapes
  outputs: List[Tensor]
  cost:    OpCost           # SymPy exprs per key cost (compute, memory, comm)
  rrt:     RRT              # resource reservation table: rows = resources,
                            #   cols = cycles, cell = units used,
                            #   e.g. ("MXU", cycle 3) -> 1
  params:  Union[InnerLoop(n_iter, body: Graph),   # loop nest
                 GraphParams(graph),               # subgraph
                 LeafParams(...)]                  # leaf op
\end{lstlisting}

An \texttt{Op} is therefore either an interior node---a loop with a trip count
and a body graph, or a plain subgraph---or a \emph{leaf}. The leaf nodes are
where the software and system sides meet: a leaf is an op for which the
\emph{system} specifies an RRT and an \texttt{OpCost}. We target
TPUs~\citep{jouppi2023tpu}, so the leaves are TPU-shaped: an MXU matmul tile
(\texttt{mxu\_op}), a VMEM tile load (\texttt{load\_tile\_to\_vmem}), or an
ICI collective (\texttt{allgather}). The algorithm side composes leaves into
loop nests; the system side prices them. Swapping either side---a new
attention variant, a new interconnect generation---touches only its own docs.

\paragraph{The builder DSL.} Models are authored in a thin Python-embedded
tracing DSL and never construct \texttt{Op} nodes by hand: a decorated block
traces into a named subgraph, decorated loops become \texttt{InnerLoop} nodes
(\texttt{@smart\_loop} is a true reduction with a carried accumulator,
\texttt{@smart\_map\_loop} a parallel map), and builder calls emit
system-priced leaves. Every dimension is a SymPy symbol, so trip counts like
$T_q/q_{\mathrm{blk}}$ stay symbolic and one trace serves the whole design
space. Listing~1 shows the (lightly condensed) flash-attention core: the
$(B,H,T_q,T_{kv})$ score matrix never leaves VMEM, and the asymmetric
$T_q$/$T_{kv}$ make the same nest serve prefill ($T_q{=}T_{kv}{=}T$) and
flash-decoding ($T_q{=}1$, $T_{kv}{=}T_{\mathrm{ctx}}$).

\begin{lstlisting}[style=dsl,float=tb,caption={},title={Listing 1: the flash-attention core in the builder DSL.}]
def flash_attention_core(b, q, k, v, *, B, H, T_q, T_kv,
                         hd_qk, hd_v, q_blk, kv_blk):
  @smart_map_loop(b, n_iter=T_q/q_blk, name="q_blocks", gather_axis=2)
  def q_block():                        # parallel MAP over query tiles
    q_i  = b.load_tile(q, (B, H, q_blk, hd_qk))
    acc0 = b.load_tile(v, (B, H, q_blk, hd_v))
    @smart_loop(b, n_iter=T_kv/kv_blk, name="kv_blocks")
    def kv_block(o_acc):                # REDUCTION over kv tiles
      k_t = b.load_tile(k, (B, H, kv_blk, hd_qk))
      v_t = b.load_tile(v, (B, H, kv_blk, hd_v))
      s   = b.mxu_op("bhte,bhse->bhts", q_i, k_t, name="scores")
      pv  = b.mxu_op("bhts,bhse->bhte", s,  v_t, name="attn_v")
      return b.tensor_add(o_acc, pv, name="accumulate")
    return b.store_tile(kv_block(acc0)) # this tile's (B,H,q_blk,hd_v)
  return q_block()                      # gathered (B, H, T_q, hd_v)
\end{lstlisting}

Distribution is expressed as \emph{sharding annotations}, not hand-placed
collectives: tensors name the mesh axes each dimension is sharded on, and a
sharded-einsum wrapper infers the collectives from the operand/output
shardings---a just-in-time \texttt{AllGather} of a sharded contracting weight,
a \texttt{ReduceScatter} when an output is reduced over a sharded dimension.
Only layout-\emph{moving} collectives are explicit: in the DeepSeekMoE block,
the dispatch \texttt{all\_to\_all} moves the expert-parallel axis from the
token-group dimension onto the expert dimension (the combine moves it back);
every other collective in the block is inferred.

\paragraph{Rolling up the Op hierarchy.} Two modes turn a tree of per-op costs
into wall-clock time. In \textbf{fast mode}, loops are rolled up coarsely:
each leaf's cost is scaled by the product of enclosing trip counts, and simple
analytical schedulers model communication/computation overlap (weight
pre-collection, all-to-all hiding) as composable transforms in the style of a
roofline bound~\citep{williams2009roofline}; evaluation is closed-form, fast enough for sweeps over thousands of design points. In
\textbf{slow mode}, each loop is modulo-scheduled~\citep{rau1994iterative}
into its resource reservation table---software-pipelining the body against
per-resource capacity---and the achieved initiation interval rolls up
recursively up the tree, yielding dependency- and resource-aware schedules for
the design points that sweeps flag as interesting.

\paragraph{Symbolic propagation.} All cost formulas are propagated upward
symbolically and every roll-up produces a closed-form SymPy~\citep{meurer2017sympy}
expression in the free variables of the design space (batch, sequence length,
bandwidths, mesh axes, datatype widths). Numeric binding happens only at the
edge---one substitution per design point---so a single symbolic build serves
an entire sweep. A design doc can even state the exact expected expression
for a collective's cost, and the generated tests assert it.

\section{Conclusion}

Developers routinely use agents to modify code. We believe the \textsc{smart} alternative approach of generating whole systems from design documentations has a lot of benefits and should increasingly be considered.
\textsc{smart} today comprises 50 design docs ($\sim$9{,}000 lines of
specification prose) spanning TPU
topology, collective cost models, numerics, schedulers, and a catalog of
frontier model families (dense, MoE, latent-attention, and robotics/VLA
variants). We regenerate with every new version change: master is reduced to
the docs plus a handful of leaf utilities, and the library is rebuilt by
sub-agent orchestration.
Treating design docs---rather than code---as the durable artifact pays the
incremental-patching debt of \Cref{eq:debt} down to zero at every
regeneration and surfaces vague intent early, because a guess must be
written into a doc to survive. The enablers---worked-example docs, a machine-discovered dependency DAG, and a minimal symbolic IR---should generalize wherever specs churn faster than software absorbs them.

\bibliographystyle{plainnat}
\bibliography{references}

\end{document}